\documentstyle[12pt]{article}

\input epsf

\begin{document}
\begin{titlepage}
\today          \hfill
\begin{center}
\hfill    OITS-678 \\

\vskip .05in

{\large \bf 
CLEO measurement of $B \rightarrow \pi^+ \pi^-$ and determination of
weak phase $\alpha$}
\footnote{This work is supported by DOE Grant DE-FG03-96ER40969.}
\vskip .15in
K. Agashe \footnote{email: agashe@oregon.uoregon.edu} and N.G. Deshpande
\footnote{email: desh@oregon.uoregon.edu}
\vskip .1in
{\em
Institute of Theoretical Science \\
University
of Oregon \\
Eugene OR 97403-5203}
\end{center}

\vskip .05in

\begin{abstract}
The CLEO collaboration has recently reported
a (first) measurement of 
$BR \left( B \rightarrow \pi^+ \pi^- \right)
= \left(
4.7^{+1.8}_{-1.5}
\pm 0.6 \right) \times 10^{-6}$. 
We study, using recent results on QCD improved 
factorization, the implications of this measurement
for the determination of 
the CKM phase $\alpha$ 
and
also for the rate for $B \rightarrow \pi^0 \pi^0$.
If the
$B \rightarrow \pi^-$
form factor is large $(\stackrel{>}{\sim} 0.3)$, then
we find that the CLEO measurement favors small $|V_{ub}/V_{cb}|$ so that
the expected error (due to neglecting the QCD penguin amplitude)
in the measurement of $\alpha$ using {\em only} the 
time-dependent analysis of the decay 
$B \rightarrow \pi^+ \pi^-$ is large $\sim 
15^{\circ}$. However, if $|V_{ub}/V_{cb}|$ is known, then it is possible
to determine the correct value of $\sin 2 \alpha$. 

\vspace{0.1in}
\end{abstract}
\end{titlepage}

\newpage
\renewcommand{\thepage}{\arabic{page}}
\setcounter{page}{1}

\section{Introduction}

Recently, the
CLEO collaboration has reported the first observation of the decay
$B \rightarrow \pi^+ \pi^-$ and a limit on the rate for $B^{\pm} 
\rightarrow \pi^{\pm} \pi^0$ \cite{poling}. In this letter, we 
determine the range of parameters, especially $|V_{ub} / V_{cb}|$ 
(entering the calculations of $B$ 
decay rates) which is preferred by this measurement/limit.
Then, we
study, in turn, the implications 
of these preferred values of parameters for the
measurement of the CKM phase
$\alpha$ using
time-dependent studies of $B \rightarrow \pi^+ \pi^-$ which will 
be done at the $e^+ e^-$ machines in the next few years and
also for the rate for $B \rightarrow \pi^0 \pi^0$. 

The effective Hamiltonian for $B$ decays is \footnote{We neglect the
electroweak penguin operators which are expected to contribute to the
$B \rightarrow \pi \pi$ decays only at the few $\%$ level.}:
\begin{eqnarray}
{\cal H}_{eff} & = & \frac{G_F}{\sqrt{2}} 
\Biggl[
V_{ub} V_{ud}^{\ast}
\left( C_1 O_1^u + C_2 O_2^u \right) 
\Biggr. 
\nonumber \\
 & & \Biggl. + V_{cb} V_{cd}^{\ast} \left( C_1 O_1^c + C_2 O_2^c \right) -
V_{tb} V_{tq}^{\ast} \sum _{i=3} ^6 C_i O_i 
\Biggr],
\end{eqnarray}
where $q = d,s$. The $C_i$'s are the Wilson coefficients (WC's) which are 
scheme- and scale-dependent; these unphysical dependences are 
cancelled by the corresponding scheme- and scale-dependences
of the matrix elements of the operators. 

In a recent paper, Beneke {\it et al.} found that the matrix elements 
for the decays $B \rightarrow \pi \pi$, in the large
$m_b$ limit, can be written as \cite{beneke}
\begin{eqnarray}
\langle \pi \pi | O_i | B \rangle & = &
\langle \pi | j_1 | B \rangle \langle \pi | j_2 | 0 \rangle \nonumber \\
 & & \times \Bigr[ 1 + \sum r_n \alpha _s^n (m_b) + O ( \Lambda _{QCD}
/ m_b ) \Bigl],
\end{eqnarray}
where $j_1$ and $j_2$ are bilinear quark currents.
If the radiative corrections in $\alpha _s$ {\em and}
$O ( \Lambda _{QCD}
/ m_b )$ corrections are neglected, then the
matrix element on the left-hand side 
factorizes into a product of a form factor and a meson decay constant
so that we recover the ``conventional'' factorization formula.
These authors computed the $O(\alpha _s)$ corrections
(in perturbation theory) using the meson light-cone
distribution amplitudes \cite{beneke}. 
In this approach, the strong interaction (final-state rescattering)
phases are included in the radiative corrections in $\alpha _s$
and thus the $O(\alpha _s)$ strong interaction
phases are determined in \cite{beneke}.
The scale- and scheme-dependence of the
WC's are cancelled by these $O(\alpha _s^n)$ corrections.

\section{Formulae for $B \rightarrow \pi \pi$}
The matrix elements for $B \rightarrow \pi \pi$ are \cite{beneke}:
\begin{eqnarray}
i {\cal M} \left( \bar{B}_d \rightarrow \pi^+
\pi^- \right) 
& = & \frac{G_F}{\sqrt{2}} \Biggl[ V_{ub} V_{ud}^{\ast} \left(
a_1 + a_4^u +
a_6^u r_{\chi} \right) \Biggr. \nonumber \\
 & & + \Biggl. 
V_{cb} V_{cd}^{\ast} \left( a_4^c +
a_6^c r_{\chi} \right) \Biggr] \times X.
\label{bpi+pi-}
\end{eqnarray}
Here
\begin{equation}
X = f_{\pi} \left( m_B^2 - m_{\pi}^2 \right)
F_0^{B \rightarrow \pi^-} \left( m_{\pi}^2 \right),
\label{X}
\end{equation}
where $f_{\pi} = 131$ MeV is the pion decay
constant and $F_0^{B \rightarrow \pi^-}$ is a ($q^2$
dependent) form factor.
\begin{eqnarray}
i {\cal M} \left( B^- \rightarrow \pi^- \pi^0 \right) & = &
\frac{G_F}{\sqrt{2}} V_{ub} V_{ud}^{\ast} (a_1 + a_2) 
\times Y,
\label{bpi+pi0}
\end{eqnarray}
where
\begin{equation}
Y = f_{\pi} \left( m_B^2 - m_{\pi}^2 \right)
F_0^{B \rightarrow \pi^0} \left( m_{\pi}^2 \right).
\end{equation}
\begin{eqnarray}
i {\cal M} \left( \bar{B}_d \rightarrow \pi^0 \pi^0 \right) & = &
\frac{G_F}{\sqrt{2}} \Biggl[ V_{ub} V_{ud}^{\ast} \left(
a_2 - a^u_4- 
a_6^u r_{\chi} \right) \Biggr. \nonumber \\
 & & - \Biggl.
V_{cb} V_{cd}^{\ast} \left( a^c_4 +
a_6^c r_{\chi} \right) \Biggr] \times 
\sqrt{2} \times Y.
\label{bpi0pi0}
\end{eqnarray}

In the above equations, the $a_i$'s are (combinations of) WC's with the
$O(\alpha _s)$ corrections added so that the $a_i$'s are scheme-
and (almost)
scale-{\em in}dependent. The values of the $a_i$'s are given in
Table \ref{a} \cite{beneke}. The imaginary parts of $a_i$'s are due to
final-state rescattering.

For the $CP$ conjugate processes, the CKM elements have to be 
complex-conjugated.

The branching ratios are given by:
\begin{equation}
BR \left( B^- \rightarrow \pi^- \pi^0 \right) = \tau _B \frac{1}{8 \pi}
| {\cal M} | ^2 \frac{|p|}{m_B^2},
\end{equation}
where $\tau _B$ is the lifetime of the $B$ meson and $|p|$ is the momentum of
the pion in the rest frame of the $B$ meson. There is a factor of $1/2$
for $\pi^0 \pi^0$ due to identical final state particles.

We neglect the
$q^2$ dependence of the form factors between
$q^2 =0$ and $q^2 = m_{\pi}^2$, {\it i.e.,} set
$F_0^{B \rightarrow \pi} \left(0 \right) = F_0^{B \rightarrow \pi}
\left( m_{\pi}^2 \right)$.
We will use two values of the form factors: $F^{B \rightarrow \pi^-} 
= 0.27$ and $0.33$ with $F^{B \rightarrow \pi^0}
= 1/ \sqrt{2} \; F^{B \rightarrow \pi^-}$. Model calculations indicate that
the $SU(3)$ breaking in the form factors is given by
$F^{B \rightarrow K^-} \approx 1.13 \; F^{B \rightarrow \pi^-}$ 
\cite{bsw,lightcone}.
The large measured 
$BR( B \rightarrow K \eta^{\prime} )$
requires $F^{B \rightarrow K^-} \stackrel{>}{\sim} 0.36$
\cite{desh} which, in turn, implies a larger value of
$F^{B \rightarrow \pi^-}$ $(\approx 0.33)$. If $F^{B \rightarrow K^-}
\stackrel{<}{\sim} 0.36$,
then we require a ``new'' mechanism to account for 
$BR( B \rightarrow K \eta^{\prime} )$: high charm content of
$\eta^{\prime}$ \cite{ali}, QCD anomaly \cite{qcd}
or new physics. Also, if 
$F^{B \rightarrow \pi^-} < 0.27$, then the 
value of $F^{B \rightarrow K}$ is too small to explain the measured
BR's for $B \rightarrow K \pi$ \cite{dutta}.

We use
$| V_{cb} | = 0.0395$, $|V_{ud}| =0.974$,
$|V_{cd} | = 0.224$, $m_B = 5.28$ GeV and $\tau _B = 1.6$ ps \cite{pdg}.

\renewcommand{\arraystretch}{1}
\begin{table}
\begin{center}
\begin{tabular}{cc} \hline
$a_1$ & $1.047 + 0.033 \; i$\\ \hline
$a_2$ & $0.061 - 0.106 \; i$\\ \hline
$a_4^u$ & $-0.030 - 0.019 \; i$\\ \hline
$a_4^c$ & $-0.038 - 0.009 \; i$\\ \hline
$a_6^{u,c} \; r_{\chi} $ & $-0.036$\\ \hline
\end{tabular}
\end{center}
\caption{The factorization coefficients for the renormalization
scale $\mu = m_b /2$ \protect\cite{beneke}.}
\label{a}
\end{table}

\section{Constraints on parameters from $B \rightarrow \pi^+ \pi^-$, 
$\Delta m_s$
and $B \rightarrow \pi^{\pm} \pi^0$}
\label{constraints}
We first comment briefly on the upper limit on $\gamma$
using the recent limit on $B^0_s - \bar{B}^0_s$ mass difference,
$\Delta m_s > 14.3 \; \hbox{ps}^{-1}$ \cite{ms}.
In the SM, we have
\begin{equation}
\frac{\Delta m_s}{\Delta m_d} = \frac{m_{B_s}}{m_{B_d}} 
\frac{B_{B_s} f^2_{B_s}}{B_{B_d} f^2_{B_d}} \frac{|V_{tb}^{\ast}
V_{ts}|^2}{|V_{tb}^{\ast}
V_{td}|^2}.
\end{equation}
With $\Delta m_d$ (the $B^0-\bar{B}^0$
mass difference) $= 0.481 \pm 0.017 \; \hbox{ps}^{-1}$ \cite{ms},
$m_{B_s} = 5.37$ GeV \cite{pdg} and
$ 
\frac{\sqrt{B_{B_s}} f_{B_s}}{\sqrt{B_{B_d}} f_{B_d}} = 1.15 \pm 0.05$ 
\cite{lattice},
we get 
\begin{equation}
\frac{|V_{td}|}{|V_{ts}|} < 0.214.
\end{equation}
In the Wolfenstein parametrization, this constrains
$| 1 - \rho - i \eta | < 0.96$ which implies
$\gamma \stackrel{<}{\sim} 90^{\circ}$. 

\begin{figure}
\centerline{\epsfxsize=1\textwidth \epsfbox{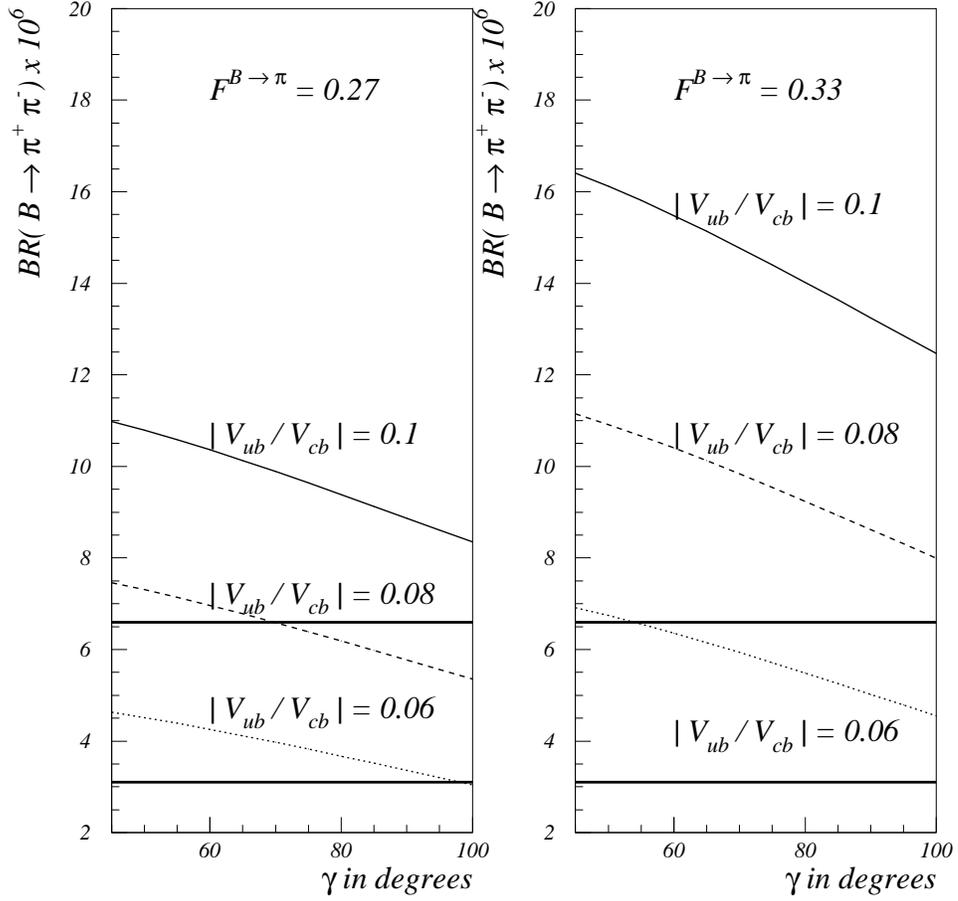}}
\caption{$CP$-averaged $BR \left( B \rightarrow \pi^+ \pi^- \right)$ 
as a function of $\gamma$ for $F^{B \rightarrow \pi^-} = 0.27$
(left) and $0.33$ (right)
and 
for
$| V_{ub} / V_{cb} | = 0.1$
(solid curves), $0.08$ (dashed curves) and $0.06$
(dotted curves). The BR measured by the CLEO 
collaboration lies (at the $1 \; \sigma$ level)
between the two horizontal (thicker) solid lines. The errors on the
CLEO measurement have been added in quadrature to compute the $1 \; \sigma$
limits.}
\protect\label{figpi+pi-}
\end{figure}

In 
Fig. \ref{figpi+pi-}
we show the $CP$-averaged BR for $B \rightarrow \pi^+ \pi^-$
as a functions of  
$\gamma$ for $F^{B \rightarrow \pi^-} = 0.33$ and $0.27$ and 
for $| V_{ub} / V_{cb} | = 0.1$, $0.08$ and $0.06$ \footnote{
The Particle Data Group quotes $| V_{ub} / V_{cb} | = 0.08 \pm 0.02$ 
\cite{pdg}. 
}. 
The CLEO measurement is $B \rightarrow \pi^+ \pi^- = \left( 
4.7^{+1.8}_{-1.5} 
\pm 0.6 \right) \times 10^{-6}$ \cite{poling}.
If 
$F^{B \rightarrow \pi^-} = 0.33$
and for $\gamma \stackrel{<}
{\sim} 90^{\circ}$,
we see from the figures that
smaller values of $| V_{ub} / V_{cb} |
\approx 0.06$ are preferred: $|V_{ub} / V_{cb}| 
=0.08$
is still allowed at
the $2 \sigma$ level for $\gamma \sim 100^{\circ}$.
However, if the smaller value of the form factor ($0.27$) is used, then
the CLEO measurement is consistent with $|V_{ub} / V_{cb}|
\approx 0.08$.
We obtain similar results using
``effective'' WC's ($C^{eff}$)'s and $N = 3$ in the earlier
factorization framework
(neglecting final state rescattering)
\cite{ali}.

\begin{figure}
\centerline{\epsfxsize=0.8\textwidth \epsfbox{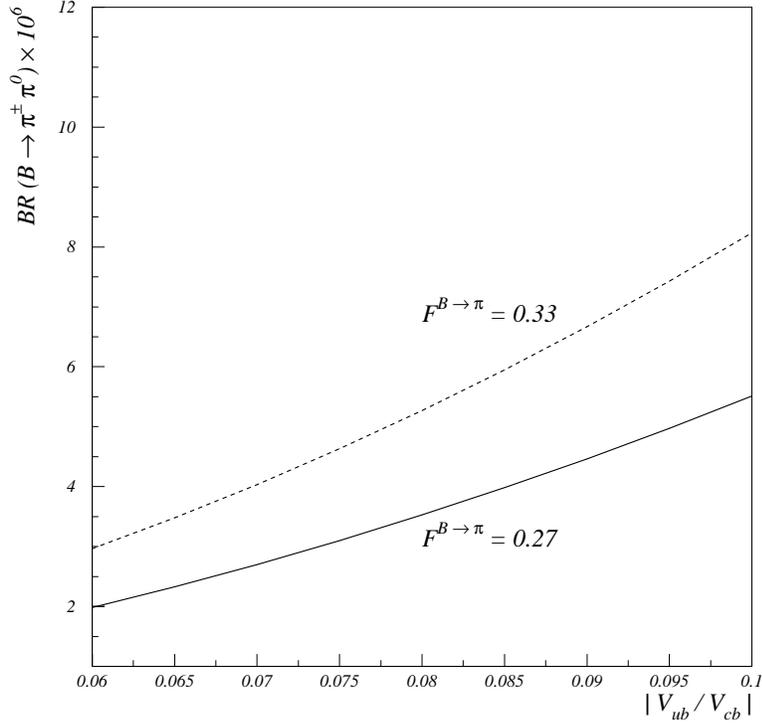}}
\vspace{-0.25in}
\caption{$BR \left( B^{\pm} \rightarrow \pi^{\pm} \pi^0 \right)$
as a function of $| V_{ub} / V_{cb} |$ for $F^{B \rightarrow \pi^-} = 
0.27$ (solid curve)
and $0.33$ (dashed curve).
The $90 \; \%$ C. L. upper limit from the CLEO collaboration is 
$12 \times 10^{-6}$.}
\protect\label{figpi-pi0}
\end{figure}

The CLEO collaboration
also quotes a ``value'' for $BR \left( B \rightarrow \pi^{\pm}
\pi^0 \right) $ of $\left( 5.4^{+2.1}_{-2.0} \pm 1.5 \right) \times
10^{-6}$, but they say that
the statistical significance of the excess over background is not sufficient
for an observation and so they quote a $90 \; \%$ C.L. upper limit of
$12 \times 10^{-6}$ \cite{poling}.
The $BR$ for $B \rightarrow \pi^{\pm} \pi^0$ is shown in Fig. \ref{figpi-pi0}
as a function of $|V_{ub} / V_{cb}|$. The upper
limit for $B \rightarrow \pi^{\pm} \pi^0$ allows
$|V_{ub} / V_{cb}|$ up to $0.1$. But, {\em assuming}
an observation at the BR quoted and if $F^{B \rightarrow
\pi-} \approx 0.33$, then there is
a preference for small $|V_{ub}/V_{cb}|$ from this decay mode
consistent with that from
$B \rightarrow \pi^+ \pi^-$.

\section{Implications for measurements of $\alpha$ and $B \rightarrow 
\pi^0 \pi^0$}
The unitarity triangle is a representation in the complex plane
of the relation: $ V_{ub} V_{ud}^{\ast} + V_{cb} V_{cd}^{\ast}
+ V_{tb} V_{td}^{\ast} = 0$. The angles of the triangle are:
$\alpha \equiv \hbox{Arg} \left( - 
V_{tb}^{\ast} V_{td} / V_{ub}^{\ast} V_{ud}
\right)$, $\beta \equiv \hbox{Arg} 
\left( - V_{cb}^{\ast} V_{cd} / V_{tb}^{\ast} V_{td}
\right)$ and $\gamma \equiv \hbox{Arg} \left( - 
V_{ub}^{\ast} V_{ud} / V_{cb}^{\ast} V_{cd} \right)$ with
$\alpha + \beta + \gamma = 180 ^{\circ}$.
Choosing $V_{cb}^{\ast} V_{cd} = - |V_{cb}^{\ast} V_{cd}|$,
we get $V_{ub}^{\ast} V_{ud} = |V_{ub}^{\ast} V_{ud}| e^{i \gamma}$
and 
\begin{equation}
\tan \beta = \frac{|V_{ub} V_{ud}| \sin \gamma}{
|V_{cb} V_{cd}| - |V_{ub} V_{ud}| \cos \gamma}.
\label{beta}
\end{equation}
Fixing
$| V_{ud} | = 0.974$ and $| V_{cd} | =0.224$ \cite{pdg}, $\beta$ 
(see above equation)
and
$\alpha$ can be obtained
as a function of $\gamma$ and $r \equiv | V_{ub} / V_{cb} |$:
\begin{equation}
\alpha = 180^{\circ}- \gamma - \tan ^{-1} 
\frac{ 
r \;
| V_{ud} | \sin \gamma}{
| V_{cd} | - 
r \; |V_{ud}| \cos \gamma}.
\label{alphatrue}
\end{equation}

The time-dependent decay rates for an initial pure $B_d$ or $\bar{B}_d$ to 
decay into a $CP$ eigenstate final state $f$ are (assuming
the total decay widths (denoted by
$\Gamma$) of the two mass eigenstates are the same):
\begin{eqnarray}
\Gamma \left( B_d (t) \rightarrow 
f \right) & = & |{\cal M}|
^2 e^{-\Gamma t}
\left( \frac{1 
+ |\lambda|^2}{2} \right.\nonumber \\
 & & \left.+ \frac{1 - |\lambda|^2}{2} \cos 
(\Delta m_d t) - \hbox{Im} \lambda \; \sin (\Delta m_d t) \right)\nonumber \\
\Gamma \left( \bar{B}_d (t) \rightarrow 
f \right) & = & 
|\bar{{\cal M}}|^2 
e^{-\Gamma t}
\left( \frac{1 + |\lambda|^2}{2} \right.\nonumber \\
 & & \left.- \frac{1 - |\lambda|^2}{2} \cos 
(\Delta m_d t) + \hbox{Im} \lambda \; \sin (\Delta m_d t) \right)
\label{timedep}
\end{eqnarray}
with
\begin{equation}
\lambda \equiv \frac{q}{p} \frac{\bar{{\cal M}}}{{\cal M}},
\end{equation}
where
\begin{equation}
{\cal M} \equiv \langle f | {\cal H}_{eff} | B_d \rangle,
\bar{{\cal M}} \equiv \langle f | {\cal H}_{eff} | \bar{B}_d \rangle
\end{equation}
and the two mass eigenstates are
\begin{equation}
|B_{L,H} \rangle = p | B_d \rangle \pm q \bar{B}_d \rangle . 
\end{equation} 
In the SM,
\begin{equation}
\frac{q}{p} = e^{- i 2 \beta}
\end{equation}
since the $B_d-\bar{B}_d$ mixing phase is $2 \beta$. 
In the case of $f = \pi^+ \pi^-$, if we neglect the 
(QCD) penguin operators,
{\it i.e.}, set 
$a_{4,6} = 0$ in Eq. (\ref{bpi+pi-}), we
get
\begin{equation}
\frac{\bar{{\cal M}}}{{\cal M}} = e^{- i 2 \gamma}
\end{equation}
and 
\begin{equation}
\hbox{Im} \lambda = \sin \left( - 2 (\beta + \gamma) \right) = \sin 2 
\alpha. 
\label{alphanop}
\end{equation}
Thus, the parameter
$\hbox{Im} \lambda$ can be obtained from the
measurement of the time-dependent asymmetry in $B \rightarrow
\rightarrow \pi^+ \pi^-$ decays (Eq. (\ref{timedep}))
and if the penguin contribution can be 
neglected, $\sin 2 \alpha$ can be determined (Eq. (\ref{alphanop})).  

In the presence of the penguin contribution, however,
$\bar{{\cal M}}/{\cal M} \neq e^{-i 2 \gamma}$ so that $\hbox{Im} \lambda \neq
\sin 2 \alpha$. We define 
\begin{equation}
\hbox{Im} \lambda = \hbox{Im} \left( e^{-i2 \beta} \frac{\bar{{\cal M}}}
{{\cal M}} 
\right) \equiv
\sin 2 \alpha_{meas.}
\label{alphameas}
\end{equation} 
as the ``measured'' value of $\sin 2 \alpha$,
{\it i.e.}, $\sin 2 \alpha _{meas.} = \sin 2 \alpha$ if the penguin 
operators
can be neglected.

\begin{figure}
\centerline{\epsfxsize=1\textwidth \epsfbox{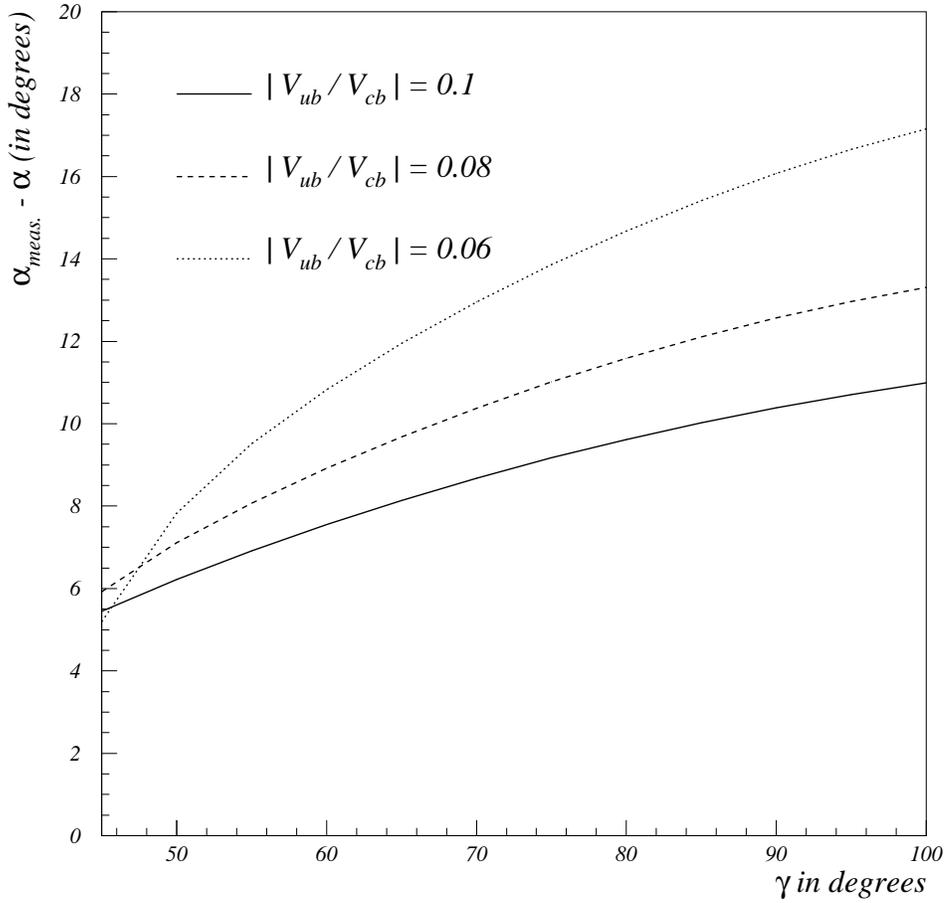}}
\caption{The error in the measurement of CKM phase $\alpha$
using (only) time-dependent $B \rightarrow \pi^+ \pi^-$ decays as a
function of 
$\gamma$
for
$| V_{ub} / V_{cb} | = 0.1$ (solid curve), $0.08$ (dashed curve) and 
$0.06$ (dotted curve).}
\protect\label{figealpha}
\end{figure}

In 
Fig. \ref{figealpha}
we plot the error in the measurement of
$\alpha$, $\Delta \alpha \equiv \alpha_{meas.} - \alpha$, where
$\alpha_{meas.}$ is obtained 
from Eq. (\ref{alphameas}) (using the amplitudes of
Eq. (\ref{bpi+pi-}) and the value of $\beta$ from Eq. (\ref{beta})) 
and
$\alpha$ is obtained from Eq. (\ref{alphatrue}). 
$\Delta \alpha$ 
depends only on $\gamma$ and $| V_{ub} / V_{cb} |$ and
is independent of 
$F^{B \rightarrow \pi^-}$ since the form factor
cancels in the ratio $\bar{{\cal M}} / {\cal M}$.
We see that for the values of 
$| V_{ub} / V_{cb} |
\approx 0.06$ preferred by the $B \rightarrow \pi^+ \pi^-$ measurement
(if $F^{B \rightarrow \pi^-} \approx 0.33$),
the error in the determination of $\alpha$ is large $\sim 15^{\circ}$ (for
$\gamma \sim 90^{\circ}$). If $F^{B \rightarrow \pi^-} \approx 0.27$, then
$| V_{ub} / V_{cb} |
\approx 0.08$ is consistent with 
the 
$B \rightarrow 
\pi^+ \pi^-$ measurement which gives $\Delta \alpha \sim 10^{\circ}$ (for
$\gamma \sim 90^{\circ}$).

\begin{figure}
\centerline{\epsfxsize=1\textwidth \epsfbox{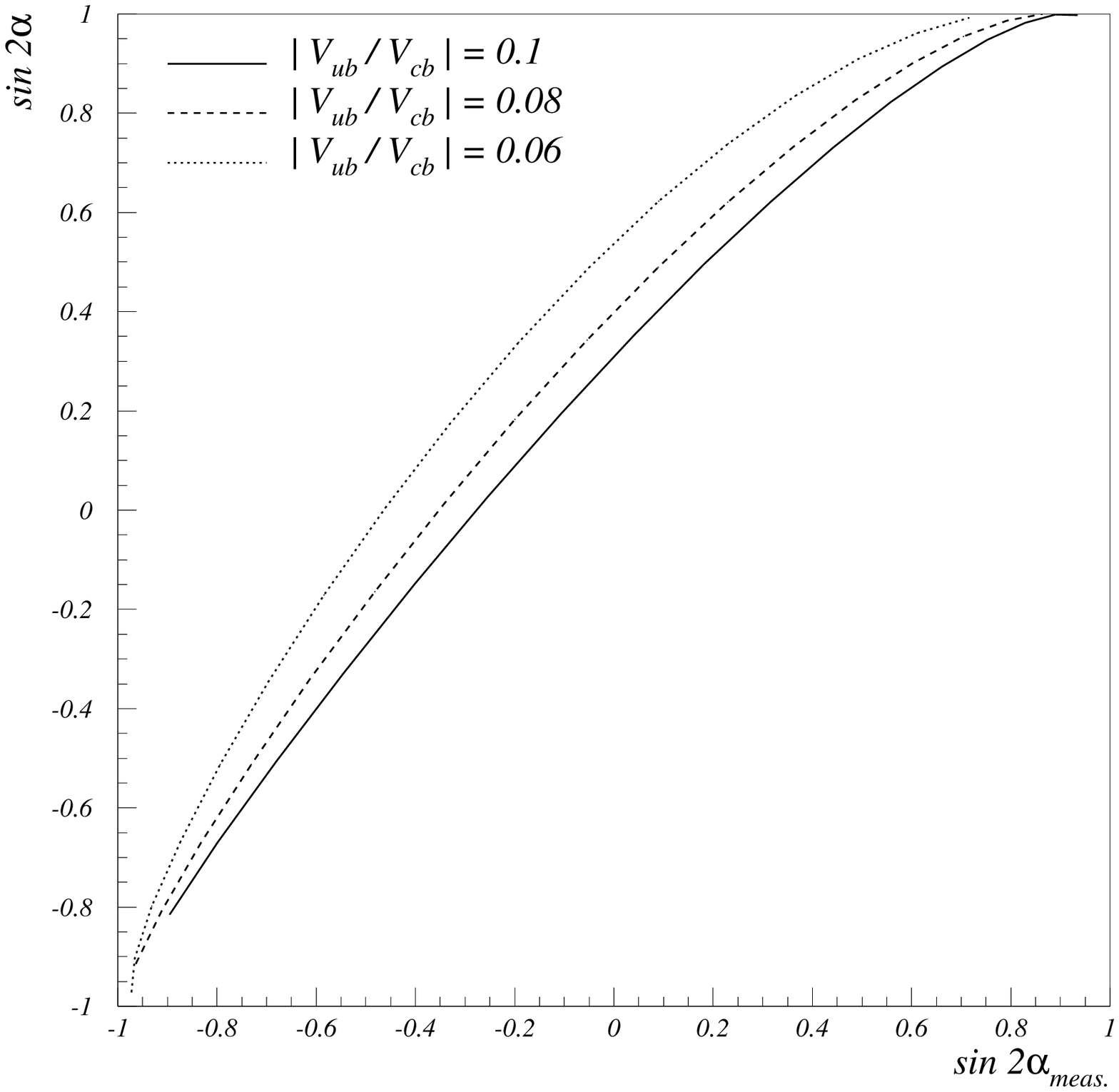}}
\caption{The ``true'' value of $\sin 2 \alpha$
as a function of the
value of $\sin 2 \alpha$ ``measured''
in $B \rightarrow \pi^+ \pi^-$ decays for $| V_{ub} / V_{cb} | = 0.1$ 
(solid curve), $0.08$ (dashed curve) and
$0.06$ (dotted curve). }
\protect\label{figrmalpha}
\end{figure}

The computation of Beneke 
{\it et al.} \cite{beneke} includes 
final state rescattering phases, {\it i.e.,} it is
{\em exact} up to $O( \Lambda _{QCD} / m_b)$ and $O (\alpha _s^2)$ 
corrections. Thus, the value of $\sin 2 \alpha$ 
``measured''
in $B \rightarrow \pi^+ \pi^-$ decays (Eq. (\ref{alphameas}))
is a known
function of $\gamma$ and $| V_{ub} / V_{cb} |$ only (in particular,
there is no dependence on the phenomenological parameter
$\xi \sim 1 / N$ and strong phases are included
unlike in the earlier factorization
framework \cite{ali}).
Since, the ``true'' value of $\alpha$ 
can also be expressed in terms of $\gamma$ and $| V_{ub} / V_{cb} |$
(Eq. (\ref{alphatrue})), we can estimate the ``true'' value of
$\sin 2 \alpha$ 
from the
``measured'' value of $\sin 2
\alpha$ for a given value of $| V_{ub} / V_{cb} |$
\footnote{In \cite{beneketalk} also a plot of
$\sin 2 \alpha_{meas.}$ as a function of $\sin 2 \alpha$ is shown,
but for a fixed $\sin 2 \beta$.}
(of course, up to $O( \Lambda _{QCD} / m_b)$ and $O (\alpha _s^2)$
corrections);
this is shown in Fig. \ref{figrmalpha}
where we have restricted
$\gamma$ to be in the range $(40^{\circ}, 120^{\circ})$
as indicated by constraints on the
unitarity triangle from present data. 
If $0^{\circ} \leq \gamma \leq 180^{\circ}$ is allowed, then there will
be a discrete ambiguity in the determination of
$\sin 2 \alpha$ from $\sin 2 \alpha _{meas.}$.
\footnote{Since 
this measurement of 
the ``true''
$\alpha$ using $B \rightarrow \pi^+ \pi^-$ decays relies on ``other''
information about the CKM matrix, {\it i.e.}, $| V_{ub} / V_{cb} |$, it is
not an ``independent'' determination of $\alpha$, but it can be used
as a consistency check.}

\begin{figure}
\centerline{\epsfxsize=1\textwidth \epsfbox{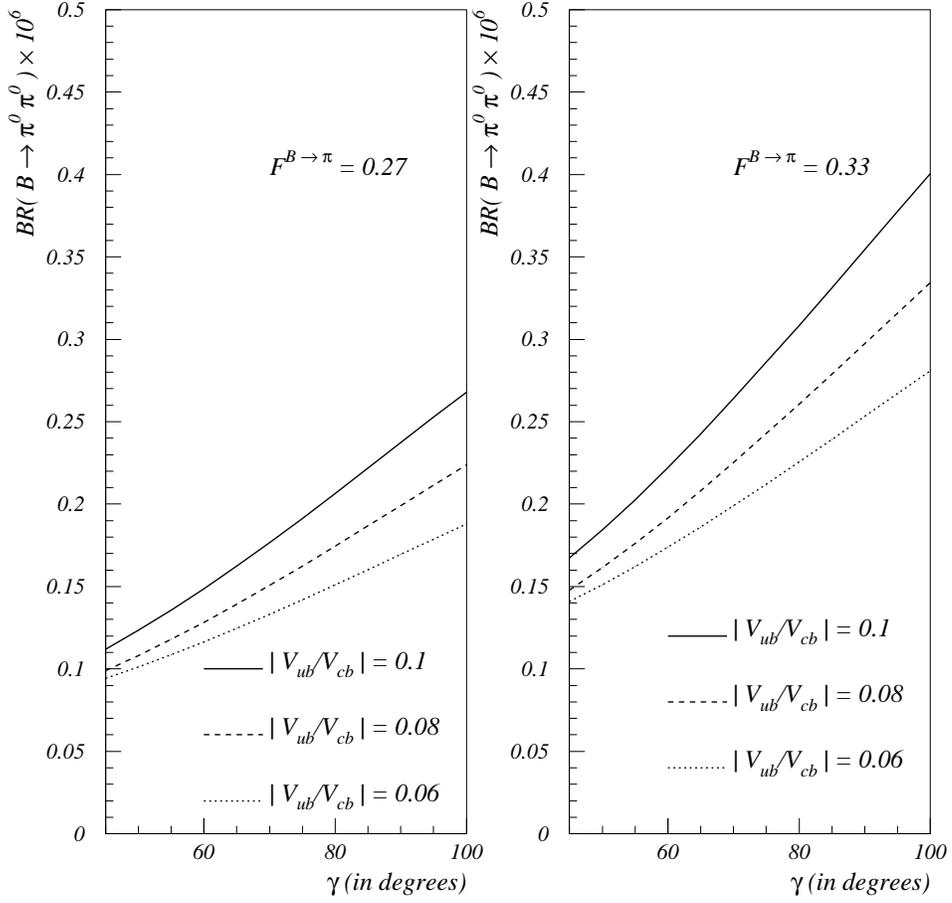}}
\caption{
$CP$-averaged $BR \left( B \rightarrow \pi^0 \pi^0 \right)$ 
as a function of $\gamma$ for 
$F^{B \rightarrow \pi^-} = 0.27$ (left) and $0.33$ (right) and
for
$| V_{ub} / V_{cb} | = 0.1$
(solid curves), $0.08$ (dashed curves) and $0.06$ (dotted curves).}
\protect\label{figpi0pi0}
\end{figure}

Gronau, London \cite{gl} showed how to
include penguin contributions in the determination
of $\alpha$, but their method 
requires, in addition to the time-dependent decay rates
for $B \rightarrow \pi^+ \pi^-$, the measurement of rates for the
(tagged) decays $B_d$, $\bar{B}_d \rightarrow
\pi^0 \pi^0$ and the rate for the decay $B \rightarrow \pi^{\pm} \pi^0$.
We show $BR \left(B \rightarrow \pi^0 \pi^0 \right)$ as a function of
$\gamma$ in 
Fig. \ref{figpi0pi0},
again for $F^{B \rightarrow \pi^-} = 0.27$ and $0.33$ and
for
$| V_{ub} / V_{cb} | = 0.1$, $0.08$ and $0.06$
\footnote{Numerically,
the coefficient $a_2$ which determines
the tree-level amplitude for $B \rightarrow \pi^0 \pi^0$
is suppressed (at lowest
order) due to a cancellation between the WC's $C_1$ and $C_2$ and thus
(unlike $a_1$) is very
sensitive to the $O( \alpha _s)$ corrections. Thus, 
we obtain slightly different
results 
using
``effective'' WC's ($C^{eff}$)'s and $N = 3$
\cite{ali}.}. We see that 
for $|V_{ub} / V_{cb}| \approx 0.06$ (which is preferred by
the $B \rightarrow \pi^+ \pi^-$ measurement for 
$F^{B \rightarrow \pi^-} \approx 0.33$), 
this rate is very small: $BR \stackrel{<}{\sim} 3 \times 10^{-7}$.
Thus, the measurement of the rates for
the (tagged) decays $B_d$, $\bar{B}_d \rightarrow
\pi^0 \pi^0$ is very difficult
in say few years of running of 
the current $e^+ e^-$ machines due to the very
small rate. Since  
time-dependent measurements of 
$B_d$, $\bar{B}_d \rightarrow \pi^+ \pi^-$ will be
achieved at these machines,
it is interesting to see how accurately
we can measure
$\alpha$ with only $B \rightarrow \pi^+ \pi^-$.

\section{Summary}
\label{sum}
To summarize, if $F^{B \rightarrow \pi^-} \approx 0.33$, then
we have shown
that the recent (and first)
CLEO measurement of 
$BR \left(
B \rightarrow \pi^+ \pi^- \right)
\left(
4.7^{+1.8}_{-1.5}
\pm 0.6 \right) \times 10^{-6}$ 
favors small $| V_{ub}/V_{cb} | \; (\approx 0.06)$.
This result is obtained using the recent computation of the
matrix elements \cite{beneke} which includes the strong interaction
phases.
The small value of
$| V_{ub}/V_{cb} |$ enhances the penguin amplitude relative to the
tree amplitude which implies that 
the error (due to neglecting the penguin contribution)
in the determination
of the CKM phase $\alpha$
using only (time-dependent) $B \rightarrow
\pi^+ \pi^-$ decays is large
$\sim 15^{\circ}$ for $\gamma \sim 90^{\circ}$. However, if
$F^{B \rightarrow \pi^-} \approx 0.27$, then $| V_{ub}/V_{cb} |
\approx 0.08$ is consistent with the 
value of
$BR \left(
B \rightarrow \pi^+ \pi^- \right)$ which implies that the error in
the determination
of $\alpha$ is $\sim 10^{\circ}$ for ($\gamma \sim 90^{\circ}$). Actually,
if $| V_{ub}/V_{cb} |$ is known, then the correct value of $
\sin 2 \alpha$ can
be determined.
Also, $| V_{ub}/V_{cb} | \approx 0.06$
implies that
$BR \left( B \rightarrow \pi^0 \pi^0 \right)$ is expected to be very
small $\stackrel{<}{\sim} 3 \times 10^{-7}$ and $BR \left(
B \rightarrow \pi^{\pm} \pi^0 \right)$ is expected to be $\sim 4 \times 
10^{-6}$.

\end{document}